\documentclass{article}
\usepackage{graphics}
\usepackage{epsf}

\def\m#1{$#1$}
\def\tr{\;{\rm tr}\;}

\newcommand{\beq}{\begin{equation}}
\newcommand{\eeq}{\end{equation}}

\newcommand{\pdr}{\partial}
\newcommand{\beqs}{\begin{eqnarray}}
\newcommand{\eeqs}{\end{eqnarray}}

\newcommand{\eps}{\epsilon}

\newtheorem{sutra}{}

\newtheorem{bhashya}{}[sutra]

\begin{document}
\bibliographystyle{h-physrev}
\input{epsf}

\title{ Yang-Mills Theory on  Loop Space\footnote{Plenary Talk at the MRST Conference 2003, in honor
  of Joseph Schechter
}}

\author{ S.G.Rajeev\thanks{rajeev@pas.rochester.edu} \\
University of Rochester. Dept of Physics and Astronomy. \\
Rochester. NY - 14627}

\maketitle
\abstract{We will describe
some mathematical ideas of K. T. Chen on calculus on loop spaces.
 They seem useful to understand non-abelian Yang--Mills theories. 
}

\section{Classical Gauge Theories }

The mathematical apparatus to describe gauge theories was discovered
almost simultaneously with the work of Yang and Mills: the theory of
connections on principal fiber bundles\cite{gaugetheorybooks}. We will now describe the three
basic examples of classical gauge theories in this language before
explaining the difficulties of formulating their quantum counterparts.

Let \m{X} be a differentiable  manifold  and \m{G} a compact Lie group. The
most interesting case is when \m{X} , which represents space-time, is
four dimensional. Also \m{G=SU(N)} is the most interesting case, the value
of \m{N} being three for quantum chromodynamics. By the imposition of
appropriate boundary conditions, we can often restrict attention to a
compact space \m{X}.

 A gauge field \m{A} is a
connection on the principal fiber bundle \m{G\to P\to X}. It is
sufficient to consider the case where \m{P} is topologically trivial,
so that it is diffeomorphic to the product \m{X\times G}:
all the physically interesting phenomena occur already in this case.

The different gauge theories are characterised by differential
equations satisfied by the connection. The simplest is
Chern-Simons-Witten theory which is the theory of flat
connections. The curvature \m{F(A)=dA+A\wedge A} is required to vanish:
\beq
dA+A\wedge A=0.
\eeq
The set of solutions of this equation ( the classical phase space) is
the same as the set of equivalence classes of representations of the
fundamental group of \m{X} in \m{G}. Thus it is a finite dimensional
space. The quantum version-essentially due to Witten \cite{csw}
 of even this simplest of all gauge theory
leads to profound new results in topology: a better understanding of
the Jones invariant of knots theory and the Witten invariants of three
manifolds. 
Note that Chern-Simons-Witten theory does not use any notion of metric
on \m{X}: it is a topological field theory. The most interesting case
is when \m{X} is three dimensional although the defining equations
make  sense in all dimensions.

The next simplest gauge theory of interest\cite{freeduhlenbeck}  is the self-dual
Yang--Mills Theory. In this case, \m{X} is a four dimensional manifold
with a Riemannian metric. The defining partial differential equation
says that half the components of curvature vanish:
\beq
F(A)=*F(A)
\eeq 
where \m{*} is the Hodge dual that maps two forms to two forms. (In
fact these equations depend only on the conformal class of the metric
tensor of \m{X}).

The deepest gauge theory of all is Yang--Mills theory, where the
connection satisfies
\beq
d_A*F(A)=0
\eeq
where \m{d_A} is the covariant derivative, \m{d_A*F(A)=d*F(A)+A\wedge
  *F(A)+[*F(A)]\wedge A}. Although other cases can be studied as toy
models,  the most  interesting case is when \m{X} is
four dimensional and the metric tensor on it is of Lorentzian
signature. Even in this case, there is by now a complete understanding of the initial
value problem \cite{moncrief}: we can regard the classical Yang--Mills theory as well-understood.

\section{ Wilson Loops}

A connection can be thought of as a one form on \m{X} valued in the Lie
algebra of \m{G}: \m{A\in \Lambda^1(X)\times \underline{G}}. This
identification depends on a choice of trivialization. A change of
trivialization is a `gauge transformation' \m{g:X\to G} which acts on
the gauge field as follows:
\beq
A\mapsto gAg^{-1}+gdg^{-1}.
\eeq
All geometrically and physically meanigful quantities must be
invariant under this transformation. Even the curvature
\m{F(A)=dA+A\wedge A} is not invariant: it transforms in the adjoint
representation:\m{F(A)\mapsto gF(A)g^{-1}}. The trace of \m{F(A)} and
its powers are gauge invariant. But they do not provide a complete set
of gauge invariant quantities from which the underlying connection can
be recovered; this is especially obvious if \m{X} is not simply connected.

The most natural gauge invariant quantities are the traces of the
holonomy of a curve. Define a function \m{S:[0,2\pi]\to G} by the
condition:
\beq
{dS(t)\over dt}+\gamma^*A(t) S(t)=0.
\eeq 
(Here \m{\gamma^*A} is the pullback of \m{A} to a one-form on the interval.) Eventhough
\m{\gamma^*A(t)} is  periodic, the solution to the above equation will
not be:  \m{U(\gamma) =S(0)^{-1}S(2\pi)} is the parallel transport (holonomy)
around the closed curve. Under a gauge transformation, \m{U(\gamma)\mapsto
  g(\gamma(0))U(\gamma)g(\gamma(0))^{-1}} so that the trace
\m{W(\gamma)={1\over N}\tr U(\gamma)} is indeed
gauge invariant. This quantity is called the `Wilson loop' in physics
jargon.

The Wilson loop is thus a complex valued function on the space of all
closed curves in \m{X}. By solving the parallel transport equation in
a power series we can get the following expansion for the Wilson loop:
\beq
W(\gamma)=\int_{\Delta_n}{1\over N}\tr
[A_{\mu_n}(\gamma(t_n))\cdots
  A_{\mu_1}(\gamma(t_1))]\dot\gamma^{\mu_n}(t_n)\cdots
\dot\gamma^{\mu_1}(t_1)dt_n\cdots dt_1.
\eeq
where \m{\Delta_n} is the simplex \m{t_1\leq t_2\leq
  \cdots t_n}.

Is it possible to reformulate the above gauge
theories in a new way where \m{W(\gamma)} is the basic variable? 

In the simplest case of Chern-Simons-Witten theory the answer is
obvious at least at the classical level. The condition of flatness of
the connections translates to the requirement that \m{W(\gamma)} be
invariant under continuous deformations of the loop: that the Wilson
loop be a function \m{W:\Omega X\to C} whose derivative is zero:
\beq
dW(\gamma)=0.
\eeq

 In
the quantum theory there will be singularities whenever \m{\gamma}
intersects itself: to get a sensible answer, \m{\gamma} must be an
embedding of the circle into \m{X}. Then \m{W(\gamma)} would be
unchanged under deformations of this embedding; i.e., a knot
invariant. In fact we can calculate the amount by which \m{W(\gamma)}
changes as \m{\gamma} is deformed through self-intersections, giving
some difference equations. The resulting difference equation for
\m{W(\gamma)}  is related to Vasiliev's approach to
knot invariants.

In the case of self-dual Yang--Mills theory, the curvature is
`half-flat'. This ought to translate to a condition that the Wilson
loop is an analytic function on the space of loops:
\beq
\bar\pdr W(\gamma)=0.
\eeq

The question arises: is there a kind of calculus on loop space with
respect to which these equations make sense? What is then the way to
rewrite classical Yang--Mills theory this way?

Only after we understand these questions we can hope to formulate and
solve quantum Yang--Mills theories this way.

\section{ Quantum Gauge Theories}

There are several ways of passing from a classical theory to quantum
theory. All of them fail to provide a mathematically well-defined quantum theory in
the case of four dimensional Yang--Mills theories due to divergences
that are characteristic of quantum field theories. There is every
reason to believe that such a theory exists however:  the other
two classes of gauge theories as well as Yang--Mills theories in
space-time dimension less than four are free of divergences. More
importantly, the profound work of 't Hooft \cite{spell}
shows that these divergences can be
removed to all orders in perturbation theory even in four dimensions:
quantum Yang--Mills theory is renormalizable. Nevertheless these
difficulties are formidable.

Loosely speaking, in the  quantum field theory, the fields are random
variables. The probability of a particular configuration is
proportional to \m{\eps{-S(A)}} where \m{S(A)} is the action. ( The
stationary points of \m{S(A)} are the solutions of the classical field
equations.) For Yang-Mills theory, for example, \m{S(A)={1\over 2\alpha}\int_X \tr
  F*F}. 

All physical quantities  follow from the expectation values of
the fields, (`correlation  functions' or Green's functions) such as 
\beq
G_{\mu_1\cdots \mu_n}(x_1,\cdots x_n)=<{1\over N}\tr A_{\mu_1}(x_1)\cdots A_{\mu_n}(x_n)>.
\eeq
We now recognize that the expectation value of the Wilson loop is a
generating function for all these Green's functions:
\beq
<W(\gamma)>=\int_{\Delta_n}G_{\mu_1\cdots \mu_n}(\gamma(t_1),\cdots \gamma(t_n)) \dot\gamma^{\mu_n}(t_n)\cdots
\dot\gamma^{\mu_1}(t_1)dt_n\cdots dt_1
\eeq
This also projects out the gauge invariant part of the correlation
functions.

The first main problem in the field (apart from the construction of
quantum Yang-Mills theory) is that this expectation value has the
asymptotic form (Wilson's Area law)
\beq
W(\gamma)\sim e^{-T{\rm Area}(\gamma)}
\eeq
for large loops. This is has been known for some time in two
space-time dimensions: an easy
result.  More recently in it has been shown (at a level of rigor
common in theoretical physics) in three space-time dimensions-a profound
physical result. It remains open even at this level in four space-time dimensions.

It would be very helpful to  rewrite gauge theories as 
field  theories  on the loop space of space-time. How does one write
differential equations in this infinite dimensional space? Surprisingly, much
of the mathematical apparatus needed for this has already developed in
the work of K. T. Chen in topology \cite{ktchen}.  In modern language, K. T. Chen
developed certain classical topological field theories on loop spaces.

\section{ K. T. Chen's Iterated  Integrals}

The set of loops on space-time is an infnite dimensional space;
calculus on such spaces is in its infancy. It is too early to have
rigorous definitions of continuity and differentiablity of such
functions. Indeed most of the work in that direction is of no value in
actually solving problems of interest (rather than in showing that the
solution exists.)

K.T. Chen's idea is to think of a function on loop space as a sequence
of functions finite dimensional spaces. The simplest example of a
function on loop space  is the
integral of a one-form on \m{X} around a curve:
\beq
\int_\gamma\omega=\int_0^{2\pi} \omega_{\mu}(\gamma(t))\dot\gamma^\mu (t)dt.
\eeq
More generally, we can have functions arising from multiple integrals like
\beq
\int_{0\leq t_1\leq t_2}\omega_{\mu_2\mu_1}(\gamma(t_2),\gamma(t_1))\dot\gamma^{\mu_2}(t_2)\dot\gamma^{\mu_1}(t_1)dt_2dt_1.
\eeq
The integrand is a tensor field \m{\omega_{\mu_1\mu_2}(x_1,x_2)}  on
\m{X\times X} which does not need to be symmetric: it is an element of
the tensor product \m{\Lambda^1(X)\otimes_C\Lambda^1(X)}. We can more
generally imagine a function on loop space as a formal power series in
such integrals:
\beq
\int_\gamma\omega:= \int_{\Delta_n}\omega_{\mu_1\cdots \mu_n}(\gamma(t_1),\cdots \gamma(t_n)) \dot\gamma^{\mu_n}(t_n)\cdots
\dot\gamma^{\mu_1}(t_1)dt_n\cdots dt_1.
\eeq
The coefficient of the \m{n}th term is an element of the \m{n}-fold product
\m{{\cal T}^n=\Lambda^1(X)\otimes_C\cdots \Lambda^1(X)}. Thus the sequence of
tensor fields \m{\omega{\mu_1\cdots \mu_n}(x_1,\cdots x_n)} can be
  thought of as defining a function on the loop space \m{LX}. There is
  a technical complication: there are certain tensor field which give
  zero upon integration a closed curve. ( For example, \m{\omega=df},
  an exact one-form on \m{X}.) Chen has identified this subspace
  \m{\cal K}. Thus we can identify the algebra of functions on \m{LX} as
  the  quotient space \m{{\cal T}/{\cal K}} where \m{{\cal
      T}=\sum_{n=0}{\cal T}^n}.

This is
  exactly the class of functions that we need to understand
  Yang--Mills theory. Notice that this fits exactly with the expansion of the Wilson loop
expectation value: the Green's functions of Yang--Mills theory are the
  coefficients. Moreover the kernel \m{K} is exactly the change in the
  Green's functions due to a gauge transformation.

There is an important change to Chen's work that is needed to apply it
to the study of quantum gauge theories. The Green's functions are
singular when a pair of points coincide (this is true even in free
field theories with linear field equatons). Thus the curves we allow
should not interesect themselves: they should be embeddings of the
circle in space-time. A function on the space of embeddings \m{EX} can be
expanded as above in iterated integrals, but the coefficients are
tensor fields on the  `configuration space' of \m{X},
\beq
F(X,n)=\{(x_1,\cdots x_n)|x_i\neq x_j \;{\rm for}\; i\neq j\}.
\eeq
These configuration spaces are interesting objects in themselves in topology.

\section{Formal Power Series in One Variable}

It is useful to look back in history to a time when calculus of one
variable itself was new to see how we should develop a calculus of an
infinite number of variables. Having understood how to differentiate
and integrate polynomials, functions were thought of as infinite
series on which similar operations couldbe defined. Even without a
theory of convergence of series ( which was developed later) it is
possible to do this in a completely rigorous way: this is the theory of formal
power series\cite{hcartan}.

We define a formal power series \m{a=(a_0,a_2,\cdots)} to be a
sequence of complex numbers, which do not need to decrease. Define the
sum product and differential  of such sequences as follows:
\beq
[a+b]_n=a_n+b_n,\quad [ab]_n=\sum_{p+q=n}a_pb_q,\quad [da]_n=(n+1)a_{n+1}.
\eeq
This can be easily verifed to be  a commutative algebra on which \m{d} is a derivation:
\beq
d(ab)=(da)b+adb.
\eeq

If all but a finite number of entries are zero, such a formal power
series defines a polynomial \m{a(z)=\sum_{n=0}a_nz^n} and the above rules are the correct rules
of adding multiplying and differentiating polynomials. We simply note
that these rules make sense even on infinite sequences even without
any convergence conditions on them. 

We can  derive similar rules for adding and multiplying functions on
loop spaces: a calculus on an infinit dimensional space can be
developed first as one on a sequence of functions each depending only
on a finite number of variables.

\section{Calculus on Loop Space}

We define a formal function   on the space of embeddings \m{EX} as a
sequence of tensors 
\beq
\omega=(\omega_0,\omega_1,\omega_2\cdots)
\eeq
where \m{\omega_n} is a  covariant tensor on  the configuration space
\m{F(X,n)}; we require these tensor fields  to be of rank one in each
variable:
\beq
\omega_n=\omega_{\mu_1,\mu_2,\cdots \mu_n}(x_1,x_2,\cdots
x_n)dx_1^{\mu_1}\otimes dx_2^{\mu_2}\cdots \otimes dx_m^{\mu_n}.
\eeq
We allow these tensor fields to have singularities as \m{x_i\to x_j},
since they only need to be well-defined on the configuration space.

It will be useful to combine the discrete label \m{\mu} and the
continuous label \m{x} into a single one \m{i} and to think of such a
tensor field  as \m{\omega_I}  where \m{I=((\mu_1,x_1),(\mu_2,x_2),\cdots
  (\mu_n,x_n)} is a sequence of discrete and continuos labels.

 By thinking of a function as a series of iterated
integrals as above (which would make perfect sense if only a finite number of
these tensors are non-zero) we can derive rules for addition (the
obvious pointwise addition will do) and
multiplication:
\beq
[\omega\circ
  \phi]_I=\sum_{S\subset\{1,2,\cdots r\}}\omega_{I_S}\phi_{I_{\bar S}}.
\eeq
This is called the shuffle product: the sum is over ways of
subdividing the  sequence \m{I} into a subsequence with labels 
in the subset \m{S}  and the  complimentary sequence labelled by
\m{\bar S}. This multiplication is obviously commutative and can be verified to be
associative; we call this the shuffle algebra \m{Sh(X)}. 
( This shuffle product is in fact the dual of the
co-multiplication in the usual tensor algebra.)

There is another,non-commutative,multiplication as well defined on
these sequences of tensors: the concatenation:
\beqs
[\omega*\phi]_{\mu_1,\mu_2,\cdots n}(x_1,x_2,\cdots
x_n)&=&\sum_{r=0}^n\omega_{\mu_1,\mu_2,\cdots \mu_r}(x_1,x_2,\cdots
x_r)\\\nonumber
& & \ \ \ \phi_{\mu_{r+1},\cdots \mu_{n}}(x_{r+1},\cdots x_n).
\eeqs

 This corresponds to the
multiplication of loops (with a common base point) where one follows
along the first loop and then the other. The two multiplication fit
into a bialgebra, indeed even a Hopf algebra. These operations  of
tensor fields on configuration spaces as well as the idea that
sequences of configuration spaces can be viewed as approximations to
spaces of embeddings are ideas originating in  algebraic topology.

The set of tensors that give zero upon integration on a curve for an
ideal \m{{\cal K}}  generated by elements of the form
\beq
u*df*v 
\eeq
where \m{f\in \Lambda^0(X)}. The shuffle algebra \m{Sh(X)/{\cal K}}
  may be thought of as a model for the algebra of functions on the
  infinite dimensional space \m{EX}.

To define a differentiation, we first define the shift operator
\m{\hat \alpha_\mu(x)}, 
\beq
[\hat\alpha_\mu(x)\omega]_{\mu_1\cdots \mu_n}(x_1,\cdots
x_n)=\omega_{\mu\mu_1\cdots \mu_n}(x,x_1,\cdots x_n).
\eeq

Then the exterior derivative is the operator
\m{\pdr_\mu\hat\alpha_\nu+\hat\alpha_\mu\hat\alpha_\nu-\mu\leftrightarrow\nu}
applied to the tensor:
\beqs
[d\omega]_{\mu\nu\mu_1\cdots \mu_n}(x,x_1,\cdots x_n)&=&
\pdr_\mu\omega_{\nu\mu_1\cdots \mu_n}(x,x_1,\cdots x_n)+\\
& & \omega_{\mu\nu\mu_1\cdots
  \mu_n}(x,x,x_1,
\cdots x_n)-\mu\leftrightarrow \nu\nonumber
\eeqs
This operation is derivation with respect to the shuffle
multiplication given above. Also, (with appropriate generalization to
exterior derivative of higher order forms) it satisfies
\m{d^2=0}. Chen uses the corresponding de Rham cohomology to derive
results in homotopy theory.

We can now check that the condition \m{dW=0} on the Wilson loop
follows from the flatness of the connnection; it is the classical
equation of motion of Chern-Simons--Witten theory translated into loop
space. Classical Yang--Mills equations also become a linear equation
on the Wilson loop \m{YW=0} where the differential operator \m{Y} is 
\beq
Y_\nu=\pdr^\mu(\pdr_\mu\hat\alpha_\nu-\pdr_\nu\hat\alpha_\mu+[\hat\alpha_\mu,\hat\alpha_\nu])+
\hat\alpha^\mu(\pdr_\mu\hat\alpha_\nu-\pdr_\nu\hat\alpha_\mu+[\hat\alpha_\mu,\hat\alpha_\nu]).
\eeq

Because of the singularities in the correlation functions, these
equations will become singular in the quantum theory. It remains a
challenge to show that these singularities can be removed: to show
that the loop equations are renormalizable, beyond perturbation theory.

The Migdal-Makeenko equations \cite{agarwalrajeev} of the large N limit of gauge theories
can now be written also as differential equations on loop space. The
framwork of Chen seems better suited to resolving singularities in
them than the original ideas of Migdal and Makeenko on Stokes
functions in removing singularities as well as solving these
equations. We hope to return to these issues elsewhere in more
detail \cite{agarwalrajeev}.

\end{document}